\documentclass[11pt]{article}
\usepackage{import}

\usepackage{epsfig,endnotes}
\usepackage[hyphens]{url}
\usepackage{hyperref}
\usepackage{caption}
\usepackage{paralist}
\usepackage{listings,multicol,filecontents}
\usepackage{float}
\usepackage{enumitem}
\usepackage[titletoc, title]{appendix}
\usepackage{datetime}
\newdateformat{monthyeardate}{%
  \monthname[\THEMONTH], \THEYEAR}
  
\graphicspath{{Tex/}}

\usepackage[a4paper,top=30mm,bottom=30mm,left=35mm,right=30mm,headsep=20pt]{geometry}
\usepackage{setspace}\onehalfspace

\begin{document}

\date{}

\title{\Large \bf Secure Containers in Android: \\the Samsung KNOX Case Study}

\author{
{\rm Uri Kanonov} \\
School of Computer Science \\
Tel Aviv University \\
\texttt{urikanonov@gmail.com}
\and
{\rm Avishai Wool} \\
School of Electrical Engineering \\
Tel Aviv University \\
\texttt{yash@eng.tau.ac.il} 
} 

\maketitle

\subsection*{Abstract}
Bring Your Own Device (BYOD) is a growing trend among enterprises, aiming to improve workers' mobility and productivity via their smartphones.
The threats and dangers posed by the smartphones to the enterprise are also ever-growing. 
Such dangers can be mitigated by running the enterprise software inside a ``secure container'' on the smartphone.
In our work we present a systematic assessment of security critical areas in design and implementation of a secure container for Android
using reverse engineering and attacker-inspired methods.
We do this through a case-study of Samsung KNOX, a real-world product deployed on millions of devices. 
Our research shows how KNOX security features work behind the scenes and lets us compare the vendor's public security claims against reality. 
Along the way we identified several design weaknesses and a few vulnerabilities that were disclosed to Samsung.

\section{Introduction}
\subsection{Background}
The wide range of possibilities provided to us by our smartphone is invaluable to our work place. Being available 24/7, 
having the ability to rapidly respond to e-mails, to open and edit documents, scheduling meetings, and attending 
video conferences regardless of our physical location, are all work-related activities. Such a setting in which the work place allows (and even encourages) 
the user to work from her personal phone is often referred to as Bring Your Own Device (BYOD). The primary reasons for supporting BYOD 
\cite{ByodAndMobileSecurity, BYODBenefitsRisksAndControlTechniques} are keeping employees satisfied (as they can use their own device), 
mobile and productive (work from anywhere). Surveys of mobile security issues 
\cite{AMultitudeOfMobileSecurityIssues, EvaluationOfSecuritySolutionsForAndroidSystems, BYODBenefitsRisksAndControlTechniques, ByodAndMobileSecurity, BYODSecurityTechnologies} 
have identified multiple problems enterprises face when dealing with BYOD. 
These include security policy enforcement, stolen or lost devices containing sensitive data, data confidentiality and 
integrity when stored on or accessed from the device. In particular the following threats are prevalent: 
\begin{description}[itemsep=-2pt,topsep=5pt,leftmargin=1em]
  \item[Untrusted Networks] 
  	In unencrypted wireless networks, attackers may eavesdrop on the communication or tamper with 
  	it by means of a man-in-the-middle (MITM) attack \cite{MobileSecurityPredictions}.
  \item[Loss or Theft] 
  	A prominent risk to mobile devices is loss or theft \cite{PhoneLossArticle1, PhoneTheftArticle1, PhoneTheftArticle2, PhoneTheftArticle3}, 
  	giving an attacker physical access to the data stored on the device. Lack of strong encryption places the data in imminent danger. 
  \item[Malware] 
  	Android has been the target for malware from the very beginning \cite{AndroidMalware1}, reaching 97\% of all mobile malware in 2014 
  	\cite{AndroidMalware2, MobileThreatReport2015}. E.g., \cite{ChineseCybercriminalsBreachedGooglePlay} describes how Chinese hackers 
  	managed to upload malicious applications to Google Play, with over a million downloads.
  	Further, in case of targeted attacks (e.g., involving industrial espionage) mobile devices may face sophisticated malware like Hacking Team's 
  	RCSAndroid \cite{HackingTeamRCSAndroid}.
  	Once malware is installed, it may be able to collect sensitive data stored on or generated by the device, 
  	for example the ability to record audio by merely using the gyroscope \cite{Gyrophone}. Such attacks usually cannot be prevented due to 
  	Android's application eco-system but may detected by anomaly detection tools \cite{AndroidKBTA, Andromaly, AppIntent}.
\end{description}

According to IDC \cite{SmartphoneOsMarketShare}, the smartphone market in 2015 Q2 is dominated by Android at 82.8\%.
Furthermore, over a quarter \cite{SmartphoneVendorMarketShare} of the Android devices sold are by Samsung. Such statistics 
warrant special attention to Samsung's Android-based devices.

Security threat mitigation for mobile devices, and BYOD-engaged devices in particular has been the target of extensive research, resulting
in a wide range of solutions. In this paper we focus on a prominent family of solutions for Android, namely ``secure containers''. A ``secure container''
is an isolated environment that provides secure storage of data, allows for confined execution of applications and controlled management of resources.  
Secure container products typically require intricate integration of trusted code often resulting in complex design and implementation.

\subsection{Related Work} \label{related-work}

There are multiple BYOD security solutions, each taking a different approach and subsequently providing different levels of security.
In this section we review representatives from each category of solutions, focusing on the advantages and shortcomings of each one.

\paragraph{Policy}
In \cite{BYODSecurityEngineering} the authors recognize that insufficient device management and security policies as well as inter-application data 
leakage are among the top failures in BYOD security. As a mitigation they propose a security framework starting from setting up a 
Mobile Device Management (MDM) system for enterprise policy enforcement, followed by device provisioning and setup and up to incident 
response in case of a detected threats. The proposed framework encapsulates the entire BYOD lifecycle, however it is in fact only a set of guidelines, 
leaving the actual implementation to the enterprise.

A concrete solution is presented in \cite{CorporateSecuritySolutionsForBYOD}. 
In this work, a ``Multi-platform Usable Endpoint Security'' system (MUSES), enforces an enterprise policy on the device. 
MUSES is a self-adapting system, utilizing data mining and machine learning to constantly refine its security policies. 
It contains various sensors such as connectivity, mail, files, location and root detection. Some key downsides of MUSES are that 
it impacts the user's primary environment and is vulnerable to root and kernel exploits. Also, in order for MUSES to be able to enforce its policy an 
application must be ``MUSES-aware'', i.e., directly ask permission from MUSES.

\paragraph{``Citrix'' Style} In \cite{NewSecurityPerspectivesAroundBYOD} the author stresses the need for an MDM as well, 
proposing an alternative solution to tackle the security issues. 
Instead of storing the enterprise data on the device, it remains on the company servers to which the employee will connect, 
operate on the data and will simply receive images of the remote screen to her device. Such an approach prevents 
loss of sensitive data in case of device loss or theft and better protects corporate data in general but it is cumbersome, vulnerable to root and kernel 
attacks and the network overhead imposed by such mode of operation may discourage employees from joining BYOD programs.

\paragraph{Root-based}
DeepDroid \cite{DeepDroid} is a custom instrumentation tool, running as the \emph{root} user, 
allowing the implementation of an enterprise fine-grained security policy in addition to Android's 
built-in permission mechanism \cite{AndroidSecurityAssessment} and the Mandatory Access Control (MAC) mechanism provided by SEAndroid.  
DeepDroid's approach has several limitations: 
(1) The \emph{root} access requirement must be satisfied by either having the vendor include it  in the firmware, or by ``rooting'' the device afterwards, 
thus exposing it to many other dangers.
(2) Relying on being \emph{root} as the ``root of trust'' opens the door for other root or kernel exploits to bypass or disable DeepDroid. 
(3) The management is done on the user's primary work environment, potentially limiting her personal (non-work) usage of the device.

\paragraph{Secure Containters}
A different approach is proposed by Cells \cite{Cells}, utilizing on-device virtualization. 
The idea is to run multiple ``virtual phones'' (VP) on a single physical phone, running each VP under its own namespace, 
isolating its applications and data from other VPs.
Resource management and isolation is achieved through special support of the kernel, which is shared among all VPs.
Cells achieves a significant degree of isolation between VPs---but not as well as can be enforced by SELinux.

Alternatively, Boxify \cite{Boxify} relies on Android's ``isolated process'' feature for application sandboxing. 
Through syscall and IPC interception it enables application separation (even between user and business domains) 
and enforcement of a security policy in a manner transparent to the sandboxed applications.

Another secure container solution is one designed by Google: ``Android for Work'' \cite{AndroidForWorkSecurityWhitePaper, AReviewOfAndroidForWork}. 
This solution utilizes the ``multiple user'' mechanism added in Android 5.0 to create a separate user for a work environment. 
This allows some isolation of the work area while maintaining the ability to share data with the personal environment.

The Achilles heel of all the solutions we mentioned so far is that a kernel 
exploit will compromise the security of the entire system. In \cite{EvaluationOfSecuritySolutionsForAndroidSystems} the authors reach the 
same conclusion regarding security solutions relying on an uncompromised kernel. 

\paragraph{Hardware-backed}
To our aid comes ARM TrustZone \cite{TrustZoneSecurityWhitepaper}, helping to move the ``root of trust" further away from the attacker. 
TrustZone is a separate environment that can run security dedicated functionality, 
parallel to the OS and separated from it by a hardware barrier. 

Utilizing the TrustZone, in DroidVault \cite{DroidVault} the authors present a security solution for applications that want to store and manipulate 
sensitive data on the device shielding it even from a compromised kernel.
This solution relies on storing the data in encrypted form on the filesystem and manipulating the unencrypted data only in the TrustZone. 
However, the limitation of DroidVault is that it is relevant only for applications that have a clear cut line between secure and insecure functionality.

Another promising solution, which is the focus of this paper, is the Samsung KNOX \cite{SamsungKNOXSecuritySolution}, 
a secure container framework built into Samsung's Android-based devices. 
KNOX, being of vendor origin, has powerful capabilities and protection from the environment (both software and hardware), 
whose ``root of trust'' is the ARM TrustZone. However, KNOX is primarily a closed-source system and its architecture is not well documented in the
open literature. One of our goals in this this work is to discover and describe the KNOX security architecture.
KNOX has been the target of some security research such as \cite{KnoxSecurityResearch2}, exposing security issues relating to password storage.
In another work \cite{KnoxSecurityResearch1}, Ben-Gurion researchers have found a vulnerability relating to security of Data in Transit (DIT) 
but have not published any technical details.

\subsection{Contributions}
In this work we present a systematic assessment of security critical areas in design and implementation of a secure container for Android 
through a case study of a real-world system which is deployed on millions of devices: 
Samsung KNOX. KNOX is a delicate combination of technologies, consisting of multiple components whose 
integration is Samsung's answer to BYOD security.
Our research, backed by extensive reverse-engineering, compares the vendor's security claims to reality, 
shows how KNOX works behind the scenes and uncovers several design weaknesses.
We also discovered some critical vulnerabilities in KNOX and presented practical attacks exploiting them. 
Our results emphasize the inherent and fundamental pitfalls in the secure container paradigm.
Finally, we contrast KNOX 1.0 with the most recent version of KNOX: we show how the latest KNOX improves security---
while also making security sacrifices in favor of user satisfaction. 

Our findings were disclosed to Samsung in December 2015 \cite{SamsungCorrespondence} and have been identified as CVE-2016-1919 
\cite{CVE-2016-1919}, CVE-2016-1920 \cite{CVE-2016-1920} and CVE-2016-3996 \cite{CVE-2016-3996}. 
In accordance with Samsung's request we delayed publicly disclosing the vulnerability described in sections 
\ref{knox-1.0-attacks-clipboard} and \ref{knox-2.x-attack-clipboard} to provide a suitable time-frame for releasing a patch. 

The rest of the paper is organized as follows. In section \ref{preliminaries} we provide some 
background material. Section \ref{knox-1.0-arch} presents our findings about the inner working on Samsung KNOX 1.0. Section 
\ref{knox-1.0-attacks} details the vulnerabilities we uncovered in KNOX 1.0 and ways to exploit them.
Section \ref{knox-2.x} presents a review of KNOX 2.3 and how it affects our attacks. Finally we present our conclusions in section \ref{conclusion}.

\section{Preliminaries} \label{preliminaries}
\subsection{ARM TrustZone} \label{arm-trustzone}
Trusted Computing has been the target of much research in the PC world \cite{TrustedComputingPlatformsTheNextSecuritySolution, OSLOImprovingTheSecurityOfTrustedComputing}
with the purpose of achieving a Trusted Computing Base (TCB), allowing the running of applications in a secure verified environment.
ARM TrustZone \cite{TrustZoneSecurityWhitepaper} is the realization of a similar concept in the mobile world through the creation of a 
Trusted Execution Environment (TEE).

ARM TrustZone is a set of security oriented extensions implemented by the processor starting with Cortex-A8, Cortex-A9 and onwards.
As stated by ARM \cite{TrustZoneSecurityWhitepaper}, TrustZone is meant to withstand software-based attacks (root exploits, kernel exploits) 
and simple (low-cost) hardware attacks, such that an attacker can attempt at home, e.g., by using a JTAG connection.

The TrustZone specification dictates that each physical processor core is separated into two ``worlds'', a \emph{normal} world and a \emph{secure} world.
The normal world is designated for running the ``Rich OS'' e.g., Android, whereas the secure world is meant to host a security-centric 
OS designed for running dedicated applications known as ``Trustlets''.
The same processor can execute both worlds in a time-sliced manner.
Each world has its own set of resources (such as flash, memory, CPU caches, etc.).
Only the secure world can access both its own and normal world resources (RAM and disk) but not vice versa.

Switching to the secure world can be done by either a dedicated assembly instruction (Secure Monitor Call - SMC) or an 
interrupt that's configured to be handled in the secure world. The secure world has full MMU support as well as its own user and privileged modes.
This allows for concurrent and isolated running of multiple trustlets.

\begin{figure*}[!h]
  \centering
    \caption{ARM TrustZone}
    \includegraphics[width=0.7\textwidth]{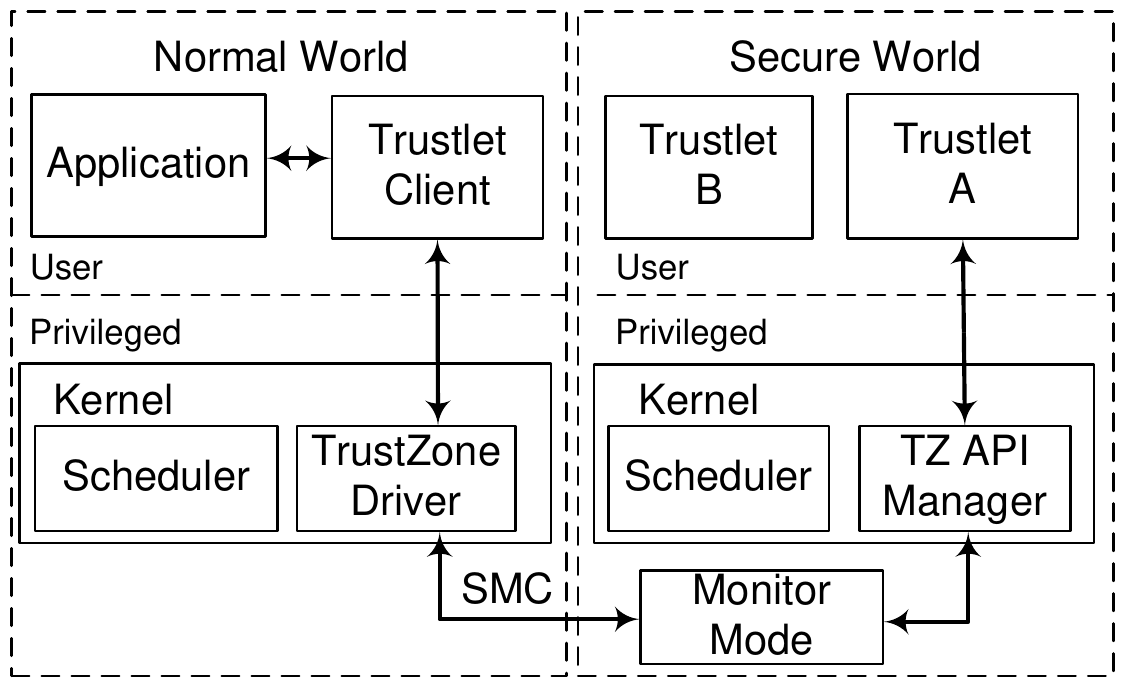}
    \label{fig:arm-trustzone}
\end{figure*}

Figure \ref{fig:arm-trustzone} depicts the basic architecture of the TrustZone secure world on the right alongside a ``Rich OS'' normal world on the left. 
In the figure we can see how a normal-world user-mode application can communicate with a Trustlet in the secure world.

Multiple proprietary secure world OS implementations exist, including QSEE (Qualcomm Secure Execution Environment) \cite{QSEE}, 
MobiCore by Giesecke \& Devrient \cite{MobiCorePresentation, MobiCoreReview} and others, all providing in principal similar services.
QSEE and MobiCore \cite{MobiCoreGalaxy3} in particular have been deployed on Samsung's Android-based devices.
These operating systems, although more compact than a traditional OS, are still vulnerable as shown by security researchers 
\cite{NextGenerationMobileRootkits}. QSEE for instance, the most widespread secure world OS implementation, has already several reported vulnerabilities 
\cite{ReflectionsOnTrustingTrustZone, QSEEExploit1, QSEEExploit2, QSEEExploit3, QSEEExploit4, HereBeDragons}. 

A similar counterpart of ARM TrustZone in the PC world is the Intel Software Guard Extensions (SGX) \cite{IntelSGXExplained}. 
SGX is a TEE for Intel processors supporting the running of trusted code in secure containers called \emph{enclaves} by 
employing dedicated hardware and processor instructions.
SGX is capable of protecting an enclave from a malicious OS or hypervisor, managing secure storage for the enclave and cryptographically attesting to its state. 
The SGX enclave is the equivalent of the TrustZone's trustlet, yet there is a key difference in the way each is run.
An enclave is run from within a regular process's runtime context in dedicated secure memory without changing the processor mode (remaining in usermode). 
On the other hand, a trustlet is run in a completely different OS, in a different context and in a dedicated processor mode.
An enclave has access to it's containing process's memory whereas a trustlet has access to the entire normal world memory.

\subsection{Samsung KNOX}
KNOX is Samsung's answer to BYOD security threats \cite{MobileMalwareAndEnterpriseSecurity}. 
Samsung announced KNOX \cite{SamsungKNOXAnnouncement} in early 2013, starting with version 1.0.0 \cite{AnOverviewOfSamsungKNOX1}. 
This version has been deployed in Android 4.3 on popular devices such as Galaxy S3 and S4. 
The most recent version is 2.3 \cite{SamsungKNOX2Announcement}, available on more advanced devices such as the Note 3 and onwards.

KNOX belongs to the ``secure container'' family, providing a secure environment alongside the user's personal environment. 
KNOX allows running enterprise applications in an isolated environment allowing to control them and configure the environment using MDM APIs.
KNOX's ``root of trust'' is its secure boot sequence, relying afterwards on runtime protections running inside the ARM TrustZone augmented by SELinux in Android.

In this section we provide highlights of the what's known about KNOX from Samsung's whitepapers \cite{SamsungKNOXSecuritySolution} 
augmented by information received from Samsung via personal communication \cite{SamsungCorrespondence}. 
In-depth analysis of KNOX 1.0 and KNOX 2.x resulting from our research is provided in sections \ref{knox-1.0-arch}, \ref{knox-1.0-attacks} and \ref{knox-2.x}.

\subsubsection{Architecture}
Samsung KNOX has a multi-tier security architecture, in which each layer is secured by its predecessor.
Looking top-to-bottom, the layers are:
\begin{description}[itemsep=-2pt,topsep=2pt,leftmargin=1em]
  \item[SEAndroid] 
  	Android itself and KNOX Applications in particular are protected by a fine-grained security policy enforced by SELinux. 
  	The policy protects applications from each other, isolates KNOX applications from user applications and partially mitigates certain attacks. 
  	Each application process has a ``context'' whose limitations (e.g., which files it can access, which processes it can communicate with) are defined 
  	by the SELinux policy. This layer's security depends on the integrity of the kernel and the security policy stored on disk.
  \item[TIMA]
  	On KNOX-enabled devices, the ARM TrustZone runs dedicated security applications whose purpose is ensuring the integrity of the Android kernel at 
  	runtime (thus ensuring SELinux's integrity). These security applications are a part of the TrustZone-based Integrity Measurement Architecture (TIMA), 
  	further described in section \ref{knox-tima}.
  \item[TrustZone]
  	TIMA relies on the protection and isolation of the TrustZone's secure world from the normal world. The protections are hardware-based 
  	(owing to the ARM processor) and software-based (narrow interface to the TrustZone, unlike the Linux kernel).
  \item[Secure Boot]
  	The initial integrity of the code running within the TrustZone and Android's Linux kernel comes from a process of secure boot, 
  	where each step in the boot chain cryptographically verifies the next step. The chain starts from the initial bootloader, fused into the ROM, 
  	being the initial root of trust. Additional details are provided in section \ref{knox-secure-boot}
\end{description}

KNOX's Trusted Computing Base (TCB) is first and foremost SEAndroid. It assumes the enforcement of the SELinux policy and the lack of a malicious kernel or root user.
All of the aforementioned security layers operate together to ensure the validity of this TCB to allow KNOX to execute in a safe environment.

In addition to Android's disk encryption, KNOX applies additional encryption to data stored within the secure container, based on a user-provided password.
To further safeguard KNOX data, sharing of data between the personal environment and KNOX container  
is blocked except for select instances like contacts and calendar events that are dependent on user configuration.

Additionally KNOX provides an extensive VPN framework intended for enterprise applications, as well as a rich set of MDM APIs allowing IT admins
an easy and effective method of managing the KNOX container.

Finally, in KNOX 1.0 the only applications that may be installed inside KNOX are those signed by Samsung, 
and downloaded from a dedicated application store. 
In contrast, in KNOX 2.x any application can be installed within the limitation of whitelists or blacklists set by each organization's IT admins. 

\subsubsection{Secure Boot} \label{knox-secure-boot}
One of the unique features of KNOX is its Secure Boot sequence.
The boot starts from the primary bootloader, burned into the ROM. It loads the secondary bootloader from flash, cryptographically verifying that it is of 
Samsung origin. The secondary bootloader verifies and loads the secure world OS. The secure world in turns runs TIMA which verifies the integrity of the 
kernel. If any of the boot components have been tampered with (e.g., by flashing a custom firmware) the device is deemed compromised, causing the 
``KNOX Warranty Bit'' \cite{KNOXWarrantyBit} to be turned on. The warranty bit is implemented via a hardware electronic fuse (eFuse) \cite{eFuse}, preventing the possibility of a rollback. 
Moreover, hashes of all loaded components as well as failures to verify components in the boot chain are stored in secure world memory to support device attestation (see TIMA section
\ref{knox-tima}).

The purpose of the ``KNOX Warranty Bit'' is not only for warranty as its name might suggest. 
If it has been turned on, the device will refuse to create and / or open any KNOX container from that point onwards. 
The rationale is that KNOX suspects the device has been tampered with and doesn't want to 
provide a potential attacker access to the user's sensitive information. 
The criterion for setting the warranty bit is the detection of irreparable alteration to the device. A primary example is flashing the device with a 
custom firmware (not signed by Samsung). Other scenarios not involving persistent modification in a way that cannot 
be isolated or recovered from (e.g., like runtime \emph{root} exploits) fall under the
responsibility of \emph{TIMA} (see section \ref{knox-tima}). We further elaborate on the subject in section \ref{knox-warranty-bit}.

An additional feature of the Secure Boot chain is \emph{dm-verity}, a hash-based verification of critical OS components. 
Its purpose is to prevent persistent rootkits from getting a foothold in the Android environment. 
\emph{dm-verity} is Linux (Android) kernel module that was introduced by Google \cite{DmVerity} starting from Android 4.4.
In stock Android \emph{dm-verity} supports only block-based verification whereas Samsung's implementation adds
additional support for file-based firmware over-the-air (FOTA) software updates.
 available stock implementation was initially introduced by Google as block
By design \cite {SamsungCorrespondence}, a failure by \emph{dm-verity} to verify a block will mark it as corrupt thereby failing any attempts to read it 
without triggering the KNOX warranty bit. 
Failing to read a critical block (e.g., due to modification of critical system binaries) will likely result in a ``soft-brick'' condition
otherwise known as a ``boot-loop''. This situation can be remedied by re-flashing a trusted firmware.
The \emph{dm-verity} feature was introduced during the lifetime of KNOX 2.x along with the Galaxy Note 4.

\subsubsection{TIMA} \label{knox-tima}
The TIMA is a major security feature of KNOX providing runtime protection. It is a set of trustlets, running within the TrustZone that provide the basis of a secure boot, 
ensure the system's integrity at runtime and provide security critical services.

\textbf{Periodic Kernel Measurement (PKM)} is a TIMA component that periodically performs validations on the kernel code and data 
(e.g., that SELinux hasn't been turned off). The precise nature of checks performed by \emph{PKM} is not documented.

\textbf{Real-time Kernel Protection (RKP)} \cite{HypervisionAcrossWorlds} is the core of KNOX's runtime security, 
guarding against kernel corruption at run-time. The main tasks of RKP include:
\begin{itemize}[itemsep=0pt, topsep=1pt] 
  \item Allowing modifications of the page tables only within the secure world, leaving them read-only in the normal world
  \item Ensuring that kernel code page are never mapped as writable
  \item Never mapping kernel data pages as executable
  \item Preventing double-mapping of kernel memory pages (in particular to user-space)
  \item Mapping all user-space memory pages as Privileged eXecute Never (PXN)
  \item Transferring control over user process credential structures to the secure world, making them read-only in the normal world
\end{itemize}
These memory defences together shield the sensitive areas that are the targets most kernel exploits to date. Kernel exploits often aim to:
\begin{inparaenum}
\item modify kernel code
\item inject code into the kernel
\item escalate process privileges
\item perform ``return-to-user'' attacks, forcing the kernel to execute user-space code
\end{inparaenum} and more; all of these should be detectable by \emph{RKP}.
Note that \emph{RKP} is still susceptible to vulnerabilities that hijack kernel execution flow and cause it to modify its own data. 
However, due to \emph{RKP}'s other defences, such vulnerabilities are much harder to turn into arbitrary code execution or full privilege escalation exploits.

The way \emph{RKP} enforces its protections on the kernel is by embedding \emph{SMC} calls 
(see TrustZone section \ref{arm-trustzone}) in key functions in the kernel code. Therefore, for example, when the kernel wishes to modify a page table entry, 
it calls a dedicated inner function, the SMC call transparently reroutes the control flow to RKP running within the TrustZone which performs the operation and returns control 
to the kernel.
This mode of operation resembles hardware-based paravirtualization \cite{EfficientVirtualization}, in which the TrustZone plays the role of the hypervisor and the kernel calls to it
to perform key tasks in a secure manner. A similar mechanism to \emph{RKP}, but with less features is presented in Sprobes \cite{Sprobes}.

Anomalies detected by both \emph{RKP} and \emph{PKM} are logged, followed by an immediate reboot of the device without setting the warranty bit. 
\emph{RKP} has been available starting from certain models of the Galaxy Note 3.
The runtime protections as well as \emph{dm-verity} are enabled if a KNOX license is present but may be enabled even without it depending on the device  \cite{SamsungCorrespondence}.

\textbf{Attestation} is an enterprise crucial feature for situations where an external application may want to validate the device's condition before 
allowing it to download sensitive data. 
The attestation is performed within the TrustZone and produces a token that can be cryptographically linked to the device indicating whether it has been 
compromised. The attestation will indicate a ``secure'' device only if it hasn't failed any security or integrity checks either while booting or 
afterwards (in particular its KNOX warranty bit must be intact). The attestation result includes the hashes of all components loaded during boot as well 
as a random nonce (against replay attacks) along with the device IMEI or Wi-Fi MAC. 

Additional important features are the \textbf{TIMA Keystore} and \textbf{SecureStorage API}, whose role is to provide secure key and data storage.

Interestingly, we can see that the TIMA's features along with the KNOX's secure boot in fact supply TPM-like functionality (Trusted Platform Module \cite{TPMSpec}) 
with several key differences. In both cases the secure boot sequence measures loaded components for the purpose of attestation.
Unlike TPM which includes all loaded software components (including OS applications) in its attestation, KNOX does not include any information 
regarding trustlets running in the secure world OS or applications running in Android (neither by the user nor by the KNOX container).
As for the TPM data sealing functionality (secure storage), it is analogous to KNOX's TIMA keystore along with the SecureStorage API.
Another key difference is that the TrustZone's functionality can always be extended by additional trustlets whereas the TPM's 
functionality is hard-wired into the chip. 

\section{The KNOX 1.0 Architecture Unveiled} \label{knox-1.0-arch}
Based on Samsung's whitepapers, KNOX looks like a very promising security solution for BYOD.
In this section we describe our own observations regarding KNOX 1.0.
We present previously unpublished discoveries regarding its design and implementation and review it from a security standpoint.  
All of our findings were uncovered through comprehensive reverse-engineering of KNOX components and are later utilized in a wide 
range of attacks on KNOX 1.0 described in section \ref{knox-1.0-attacks}.
Complementary analysis of KNOX 2.x is presented in section \ref{knox-2.x}.

\subsection{Research Environment}
Our research combined both static and dynamic analysis of KNOX 1.0.0 on two Samsung devices running
Android 4.3:
\begin{description} \setlength{\itemsep}{0pt}
  \item [Galaxy S3] \hfill \\ Model GT-I9305 \\ Kernel version 3.0.31 \\ Build Number: JSS15J.I9305XXUEML8
  \item [Galaxy S4] \hfill \\ Model GT-I9505 \\ Kernel version 3.4.0 \\ Build Number: XXUEMJ5.CCOM
\end{description}

Given that KNOX is closed-source, our static analysis had to be performed by pulling relevant binaries from the device 
and reverse-engineering them. We disassembled native libraries using ``IDA Pro'' \cite{IDAPro}. 
As for the Java in Android, the byte-code comes in the form of Dalvik Executables (.dex), run by the Android's Java runtime engine: the Dalvik VM.
We often encountered .odex (optimized dex) files, which we converted to .dex using ``Universal Deodexer'' \cite{UniversalDeodexer}. 
The .dex files we converted to .jar files using ``dex2jar'' \cite{dex2jar} and finally disassembled to Java code using ``jd-gui'' \cite{JDGui}.

Performing dynamic analysis (runtime modification of code and placement of hooks) required root privileges in order to bypass Android and 
KNOX's security features.
The Galaxy S3 in our possession was initially rooted by flashing a custom firmware, setting its warranty bit, making KNOX unusable.
Indeed when attempting to run KNOX we encountered an error stating ``Your device is not authorized to enter Samsung KNOX mode". 
This drew us to search for an alternative device to experiment on, the Galaxy S4. The S4 was rooted by a personally modified version of \emph{SafeRoot} 
\cite{SafeRoot}, which utilizes a kernel exploit from CVE-2013-6282 \cite{CVE-2013-6282} for privilege escalation to root. 
This method was not detected by KNOX, leaving KNOX fully functional and allowing us to run alongside it as root. Our modification to the basic \emph{SafeRoot} 
included removing functionality in charge of disabling KNOX. This was necessary to prevent the rooting process from tampering with our research environment. 

Once we had root access without tripping the KNOX warranty bit, our dynamic analysis was performed using \emph{Xposed} \cite{Xposed}, 
a framework allowing the modification of Java code at runtime. 
Moreover, as we advanced in our research, we leveraged \emph{Xposed} to blind KNOX on the Galaxy S3, returning it to full functionality 
while retaining our root privileges (more on this in section \ref{knox-1.0-hide-warranty-bit}).

\subsection{Application Architecture}

\begin{figure*}
  \centering
    \caption{The KNOX Architecture}
    \includegraphics[width=\textwidth]{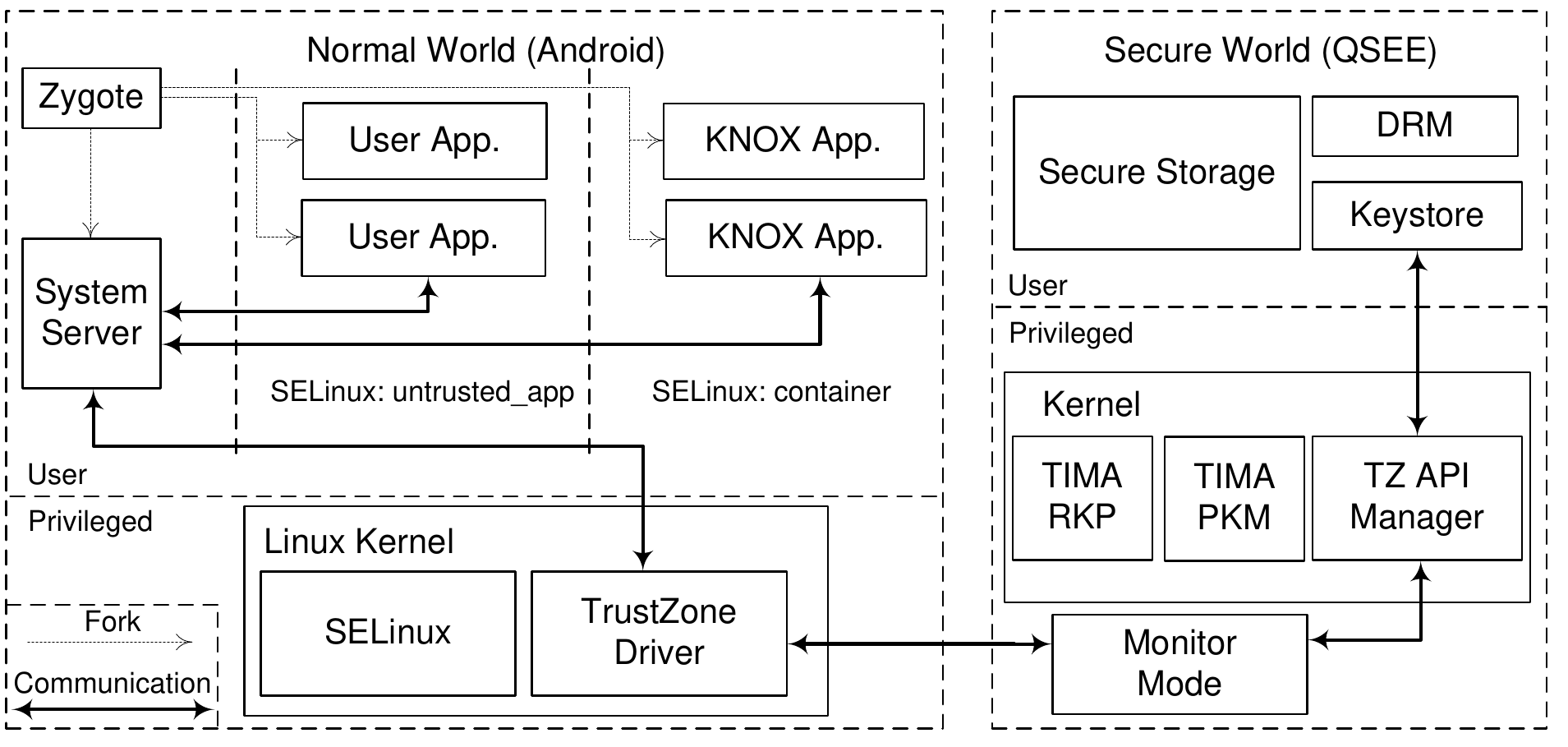}
    \label{fig:knox-architecture}
\end{figure*}

\subsubsection{The KNOX Architecture} \label{knox-1.0-architecture}
The KNOX whitepapers create the impression that KNOX applications run in a separate ``container'' having nothing in common with 
regular applications, but we discovered that things are somewhat different. 
Let us recap how normal applications are run in Android. On boot, the \emph{init} process starts \emph{Zygote}, the initial Dalvik VM. 
\emph{Zygote} in turn forks and runs the \emph{system\textunderscore server} which hosts all the OS-provided Java services. Running applications is done 
by communicating with \emph{Zygote} and asking it to fork, switch to the appropriate SELinux context and run the requested application's entry point. 
This design makes the \emph{Zygote} process the parent of all application processes and was conceived in order to maximize shared code between 
applications (by loading common code in \emph{Zygote}).
KNOX applications are run precisely in the same manner as normal applications: they are forked from \emph{Zygote} but under a 
dedicated SELinux context named ``container''. 
This context isolates the KNOX applications from other user applications and even hides them when enumerating processes. 

In our opinion, relying on SELinux for process isolation is a reasonable design choice. However it does leave a gap: if \emph{Zygote} is compromised, 
for instance if an attacker manages to run malicious code inside it, the attacker's code would propagate to all KNOX applications and the 
\emph{system\textunderscore server}: in fact, this is how \emph{Xposed} works. If SELinux protection is bypassed by some vulnerability, 
none of KNOX's defence mechanisms are equipped to detect or defend against Java code injection. We leverage this security gap in several of our attacks 
(\ref{knox-1.0-keyboard-sniffing}, \ref{knox-1.0-screen-capture}, \ref{knox-1.0-hide-warranty-bit}).

Running KNOX is done through a dedicated application named the ``KNOX Container Agent'', mostly responsible for the login UI.
Once the password is entered, the container agent asks the Enterprise Container service (\emph{container\textunderscore service}) to validate it. 
Upon successful validation the container agent requests the \emph{container\textunderscore service} to mount the encrypted filesystem, after which it runs 
the KNOX home screen application. Figure \ref{fig:knox-architecture} shows the architecture of KNOX 1.0 that we discovered. 

\subsubsection{Application Wrapping} \label{knox-1.0-app-wrapping}
In KNOX 1.0 the user cannot simply install any application she chooses into KNOX; she must download a previously approved
``wrapped" application (signed by Samsung) from a dedicated KNOX app store.
In addition to adding a procedure of review and validation for applications, application wrapping is necessary due to a technological limitation.
Each application in Android is uniquely identified by its ``package''(e.g., com.android.email is the Email application), therefore one cannot install 
two different applications with the same package. Since KNOX applications run alongside to the normal applications under 
the same ``application namespace'', this means that there can't be both ``user'' and KNOX instances of Email installed on the same device. 
The solution Samsung selected is wrapping: they repackage the application under a modified package name by prefixing 
\emph{sec\textunderscore container\textunderscore 1.} (e.g., sec\textunderscore container\textunderscore 1.com.android.email).
Application wrapping makes it more difficult for developers to develop applications for KNOX 1.0 but on the other hand, it creates a barrier 
making it more difficult for malware to find its way into KNOX. 
As a side note, this limitation prevented us from legitimately running our own ``applications'' under KNOX 1.0.

\subsubsection{Shared Services}
Another implication of the ``side-by-side'' design, is sharing the services between KNOX and the user applications through the 
\emph{system\textunderscore server}. 
This central process hosts generic services such as \emph{input\textunderscore method} (keyboard), \emph{clipboard} and \emph{connectivity} 
in addition to KNOX-specific services such as \emph{tima} (TIMA service) and \emph{container\textunderscore service} (Enterprise Container service). 
Both user and KNOX applications can communicate with \emph{all} of these services and it is the services' responsibility to verify that their 
client has sufficient privileges before performing any action or relinquishing information.
We take advantage of this observation in the two attacks in sections \ref{knox-1.0-attacks-clipboard} and \ref{knox-1.0-attacks-vpn-mitm}.

\subsubsection{Lessons Learned}
\paragraph{Component Reuse}
Running KNOX side-by-side with the regular Android applications, separating the processes using \emph{SELinux} is a reasonable design choice given that 
it reuses an existing security feature of the operating system. The reuse of an already thoroughly tested and widely deployed security mechanism is much
less risky than the invention of a new proprietary security mechanism with a high potential of undiscovered security vulnerabilities.
On the other hand, reuse of components, such as the \emph{system\textunderscore server} as a cornerstone of KNOX exposes parts of the 
secure container's critical infrastructure to potential attackers.
To conclude, component reuse is welcome, given proper protection for the added attack surface.

\paragraph{Managed Code Protection}
Given that KNOX utilizes managed (Java) code as a part of its security infrastructure (\emph{system\textunderscore server}) and allows to run 
Java-based applications, it must adequately protect the managed-code layer. 
Otherwise, without proper protection the secure container becomes highly vulnerable to code-injection attacks, as we consistently show in our attacks (section \ref{knox-1.0-root-attacks}).

\paragraph{Application Validation}
Although \emph{Application wrapping} exists in KNOX 1.0 only due a technical limitation it is an excellent security feature. As already proven by iOS, 
compared to Android, a thorough process of application review by a trusted party, greatly diminishes the risk of a malware making it to the 
application store. The same benefit can be drawn from Samsung being the sole authority for allowing an application to be installed within KNOX 1.0.

\subsection{Data Encryption and Password}
One of KNOX's primary features is On-device Data Encryption (ODE), whose role is to protect the sensitive corporate data stored within the KNOX container.
We discovered that KNOX implements ODE using two filesystem partitions, one for the application data (/data) and one for the sdcard (/sdcard).
Each of the two partitions is mounted as an eCryptFS encrypted file system \cite{eCryptFS}. The secret on which the entire data encryption scheme relies 
is the user's password.

\subsubsection{eCryptFS} \label{knox-1.0-ecryptfs}
The eCryptFS filesystem is file-based, contrary to the stock full disk encryption used by Android (via \emph{dm-crypt}). Additional, Samsung also uses 
eCryptFS for sdcard encryption.
The encrypted application data is stored in \emph{/data/.container\textunderscore 1} and is mounted to \emph{/data/data1} whereas the encrypted 
sdcard is stored in \emph{/storage/container/.sdcontainer\textunderscore 1} and mounted to \emph{/mnt\textunderscore 1/sdcard\textunderscore 1}. 

The input for eCryptFS is a password 4 to 32 bytes long. In usage scenarios like Ubuntu Home Directory encryption, 
the encryption password is directly the password provided by the user.
However, KNOX uses a more elaborate scheme. In KNOX the encryption password, called the ``eCryptFs Key''  is a combination of the user's password 
(minimum 7 chars) and 32 random bytes (denoted by \emph{TIMA key}). 
We shall discuss how this eCryptFS key is generated in section \ref{knox-1.0-ecryptfs-key}.
To allow the user to change the password without re-encrypting the data, eCryptFS uses a \emph{Data Encryption Key (DEK)} and a \emph{Master Key (MK)} 
which encrypts the DEK. The MK is derived from the password (eCryptFS Key) and a salt using PBKDF (Password-based Key Derivation Function) \cite{PBKDF}. 
The \emph{Encrypted DEK (EDK)} along with the salt (for the MK) and an HMAC on the EDK is named the \emph{EDK Payload}. 
In KNOX the EDK Payload is stored in \emph{/data/system/edk\textunderscore p\textunderscore container\textunderscore 1}, accessible only by root.
This encryption scheme is as described in \cite{SamsungKeyManagement} and has been verified by us via reverse-engineering the relevant binaries.

In Android the filesystem management falls under the responsibility of \emph{vold}, a \emph{root} daemon receiving its commands from a 
UNIX-domain socket owned by \emph{root} and \emph{mount} (limiting its attack surface).
As we recall, the mounting of the data and sdcard containers is done by communicating with the \emph{container\textunderscore service}. 
The service in turn (after going through several middlemen), conveys the mount request to \emph{vold} in textual form.
\emph{Vold}'s functionality is split across shared objects (Linux dynamically loaded libraries). 
The mount request is handled by \emph{libsec\textunderscore ecryptfs.so}. It reads the EDK payload using \emph{ssRead} from the aforementioned path, 
validates it (using the HMAC) and decrypts it obtaining the DEK. In section \ref{knox-1.0-dek-attack} we outline attacks aimed at obtaining the DEK.

\subsubsection{TIMA Key} \label{knox-1.0-tima-key}
As we mentioned in section \ref{knox-1.0-ecryptfs}, a part of the creation of \emph{eCryptFS key} is the \emph{TIMA key}.
The TIMA key, a sequence of 32 random bytes, is generated by the Enterprise Container service during the first KNOX container creation. 
The user can delete and recreate the container afterwards but the TIMA key will remain the same.
Once generated the Enterprise Container service asks the TIMA service to ``install'' the key.
When requested to mount the encrypted file system, the Enterprise Container service ``retrieves'' the key from the TIMA service, combines it with
the password and passes the mount command forward to \emph{vold}.
The \emph{install} and \emph{retrieve} operations are forwarded to the TIMA Keystore trustlet running in the ARM TrustZone. 
The TIMA service will provide the key to any thread running within the \emph{system\textunderscore server} or process with \emph{system} UID.
In the Galaxy S3 the TIMA Keystore trustlet is run by MobiCore whereas in the Galaxy S4 it is run by QSEE. 

In our opinion, a weakness in this procedure is that the TIMA key should have been generated within TrustZone and not in the normal world Android.
Not doing so makes the TIMA key available in normal world memory, exposing it to an attacker with root privileges. 
This design weakens the concept of a TrustZone-supported environment, where the TIMA keystore is expected to keep the 
keys safe in the secure world. 
We leveraged this weakness along with the observations from section \ref{knox-1.0-architecture} 
to obtain the TIMA key using \emph{Xposed}. 

\subsubsection{Inputting the User Password} \label{input-user-password}
The ARM TrustZone supports secure I/O which does not pass through the normal world as demonstrated by DroidVault \cite{DroidVault}. 
However, KNOX does not leverage this possibility for the input of the user's password.
This design means that a transient copy of the user's password resides in the memory of the following processes:
\begin{enumerate}[itemsep=0pt]
\item The KNOX Container Agent, which presents the password textbox
\item The on-screen keyboard process that handles the touchscreen input
\item The \emph{system\textunderscore server} that receives the password via IPC and uses it to create the eCryptFS key
\end{enumerate}

The only protection around the password that Samsung have put in place is that KNOX only agrees to use the Samsung official keyboard and no other.
This is indeed a prudent choice as a malicious keyboard can capture the user input, but other intervention points exist.
An attacker with root privileges could intercept the password in each of the aforementioned processes. 
Case in point, using \emph{Xposed}, we exploited this weakness in order to obtain the user's password from within both the KNOX Container Agent
and the \emph{system\textunderscore server}. This only strengthens our observation regarding necessaity of protection of Java code 
(see section \ref{knox-1.0-architecture}).

\subsubsection{Verifying the User Password}
Once the user enters the password and attempts to login into KNOX, the Enterprise Container service ``verifies'' the password.
An improper validation scheme can compromise the password and make an attacker's life much easier \cite{DoYouThinkYourPasswordsAreSecure}.
We found that upon container creation (first password entry), the Enterprise Container service calculates a salted hash of the password and stores it in 
\emph{/data/system/container/containerpassword\textunderscore 1.key} (owned by \emph{system}). The salt is stored in the ``secure" system
settings which are read-only accessible to all. The hashing algorithm is described in Listing \ref{lst:HashAlgo} and uses 1024 activations of SHA1.

To obtain the hash an attacker requires \emph{system} or \emph{root} privileges, and even so it is difficult to brute-force.
However, an artifact found in the code, shown in Listing \ref{lst:OriginalHashAlgo}, 
indicates that there was a previous weaker hashing scheme, possibly in earlier versions of KNOX, which only concatenates MD5 and SHA1 hashes. 
Whenever verifying the password, the Enterprise Container service decides according to the hash length which scheme to compare against.
This alternative scheme is much weaker than the current implementation in terms of brute-force complexity. 
Nevertheless, in our opinion both methods are reasonably secure.

\begin{lstlisting}[caption={Current Hash Algorithm}, label={lst:HashAlgo}]
function hash(password, salt) {
  saltedPass = password + salt;
  sha1 = new SHA1();
  digest = null;
  for (int i = 0; i < 1024; ++i) {
    if (digest != null) {
      sha1.update(digest);
    }
    sha1.update(i);
    sha1.update(saltedPass);
    digest = sha1.digest();
  }
  return toHex(digest);
}
\end{lstlisting}

\begin{lstlisting}[caption={Previous Hash Algorithm}, label={lst:OriginalHashAlgo}]
function hashOriginal(password, salt) {
  saltedPass = password + salt;
  sha1Hash = SHA1.digest(saltedPass);
  md5Hash = MD5.digest(saltedPass);
  return toHex(sha1Hash) + toHex(md5Hash)
}
\end{lstlisting}

\subsubsection{Lessons Learned}
\paragraph{Using the TrustZone} \label{knox-1.0-data-encryption-lesson}
KNOX 1.0 has a complex data encryption scheme, with many shortcomings and even crucial vulnerabilities. 
A likely reason for this particular design, especially in KNOX 1.0 is Samsung's desire to provide data encryption capabilities with maximal component 
reuse and minimal ``glue'' code. Let us look at some of the reused components:
\begin{description}[labelindent=1cm] \setlength{\itemsep}{0pt}
  \item [system\textunderscore server] Already existing and easily extensible with additional services for data encryption (exposes the user's password and TIMA key).
  \item [vold] Responsible for mounting and unmounting filesystems in Android, forcing it to be the next stop after the \emph{system\textunderscore server} (exposes the DEK)
  \item [eCryptFS] Already used by Samsung for sdcard encryption, only needing to combine it with KNOX (exposes the encrypted FS contents).
\end{description}
The desire to reuse the above components led to an insecure design leaving both critical key components (user password, TIMA key and DEK) 
and the user's data in plaintext in Android. 
 
The main flaws in Samsung's design come from not using the TrustZone enough. We suggest the following design, widely utilizing the TrustZone 
as the secure container's strongest ally.
\begin{description}[labelindent=1cm]
\item[Password Input] Reading the password from the UI only from within the TrustZone, never exposing it to Android.
\item[Key Derivation] Using a well known key-derivation algorithm directly (like PBKDF2) to avoid vulnerabilities like the one described in section \ref{knox-1.0-ecryptfs-key} 
from within the TrustZone to avoid exposing the password and encryption key (DEK) to Android, only providing Android a handle to the key.
\item[Data Access] The data encryption and decryption must occur within the TrustZone. Querying and modifying the data should be allowed only through a 
controlled interface using the TrustZone driver and only given the correct key handle.
\end{description}
The clear advantage of such a design is critical information such as encryption keys or the entire decrypted filesystem never leaves the safety of the
TrustZone.

\paragraph{Password Verification}
KNOX uses an adequate password verification scheme, avoiding persistently storing the password in plaintext and reasonably mitigating brute-force attacks.
It is a good design for any secure container implementation.

\subsection{KNOX Warranty Bit and TIMA} \label{knox-1.0-tima}
\subsubsection{The Need for TIMA}
The TIMA is a crucial building block in the security structure of Samsung KNOX as is responsible for continued preservation 
of the system's integrity after boot and maintaining the TCB.
The lack of full functionality of TIMA on the Galaxy S4 (\emph{RKP} and \emph{dm-verity} in particular) allowed us to root the device and 
install the \emph{Xposed} framework, both necessary for our research. This however, only stresses the importance of the TIMA runtime protections. 
For instance, if \emph{RKP} had been deployed and the process credential structure protection had been enabled 
(see section \ref{knox-tima}), we would not have been able to root the device using \emph{SafeRoot}. 
This is due to the fact that \emph{SafeRoot} elevates the running process' privileges by modifying the credential structure.
Additionally, \emph{dm-verity} would have flagged our replacement of \emph{Zygote}'s binary in the system image, causing the device to enter 
a ``boot-loop''.

\subsubsection{The KNOX Warranty Bit} \label{knox-warranty-bit}
Recall that the warranty bit is the indicator whether a device has been compromised or not. 
It is the TIMA's responsibility to fail KNOX container creation, and any access to an existing container on a device, if the warranty bit has been set.
This functionality is present starting from the Galaxy S3 as indicated by the fact that our rooted Galaxy S3 refused to create a KNOX container.
By tracking the container creation process, we saw that the API that fails was the TIMA key installation via the TIMA service. 
This shows that TrustZone TIMA keystore's functionality does depend on the warranty bit. In section \ref{knox-1.0-hide-warranty-bit} we show 
how a root attacker can overcome this difficulty quite easily.

\subsubsection{Two TrustZones?}
Android's communication with the TIMA trustlets (running in the TrustZone) goes through a dedicated driver of the secure world OS running in Android
as depicted in Figure \ref{fig:knox-architecture}.
Communication with trustlets running in the MobiCore OS is done through \emph{/dev/mobicore} and with QSEE trustlets through \emph{/dev/qseecom}.
We found that whereas on the Galaxy S3 only the MobiCore device exists, on the Galaxy S4 both of them do.
We observe that the TIMA keystore is implemented as a MobiCore trustlet on the S3 and as a QSEE trustlet on the S4.
On the other hand, the SecureStorage API is implemented on both devices as MobiCore trustlet.
Given that both MobiCore and QSEE are full ``Secure World'' operating systems on their own, the fact that they manage to co-exist
leads us to believe that Samsung had to work quite hard to make sure they don't step on each other's toes. 
Such forced co-existence can often lead to security problems and definitely warrants a closer inspection. We did not pursue this direction further.

\subsubsection{Lessons Learned}
\paragraph{Hardware Root of Trust}
A sound design decision made by Samsung is the layered security model where each security layer is protected by its predecessor, drawing the initial
trust from the hardware. This design, if implemented properly can provide the highest level of security for both the KNOX infrastructure and its applications.
A critical link in the chain is the TIMA which protects the device from various root or kernel exploits and maintains the TCB. Owing the hardware root-of-trust,
the secure container can safely rely on the TIMA protection. We recommend any secure container implementation to base its TCB on hardware support, e.g., like the ARM TrustZone.

\subsubsection{Denial of Service Attacks}
An additional concern with TIMA is that it may actually work all too well. As suggested to us by Y. Elovici \cite{TIMADosAttack}, 
this may lead to a possible Denial of Service (DoS) attack. E.g.,
an attacker could acquire exploits that are capable of remounting the \emph{/system} partition, thereby tripping \emph{dm-verity} 
and ``soft-bricking'' the device.

\section{Attacks on KNOX 1.0} \label{knox-1.0-attacks}
In this section we describe several vulnerabilities we found in KNOX 1.0 and attacks that exploit them.
Complementary attacks on KNOX 2.x are described in section \ref{knox-2.x-attacks}.

We divide the attacks into two categories, those requiring \emph{root} privileges 
(i.e., needing an exploitable vulnerability leading to privilege escalation) and those that don't. 
The vulnerabilities below have been communicated to Samsung in December 2015 \cite{SamsungCorrespondence} and 
published on \emph{bugtraq} \cite{bugtraq} in January 2016 and April 2016 as CVEs \cite{CVE-2016-1919, CVE-2016-1920, CVE-2016-3996}.

\subsection{Root-not-required Attacks} 
The attacks presented in this section do not violate KNOX's TCB as they do not require any root privileges and aren't hindered by SEAndroid.

\subsubsection{VPN Man-in-the-Middle} \label{knox-1.0-attacks-vpn-mitm}
In this section we present CVE-2016-1920 \cite{CVE-2016-1920}, a vulnerability which allows a user application running outside KNOX to perform a Man-in-the-Middle (MITM) attack against KNOX SSL/TLS traffic.
The vulnerability is the combined result of the following facts:
\begin{enumerate}[itemsep=0pt]
\item Android manages X.509 certificates through a certificate store. This store is used when validating SSL/TLS certificates using the chain-of-trust scheme.
\item Android supports installing 3rd party certificates (involving user interaction).
\item Crucially, a side-effect of the service sharing in the KNOX 1.0 design is that the same certificate store applies to both Android and KNOX applications.
	This means that when validating server certificates in SSL/TLS connections, KNOX applications relying on chain-of-trust verification will trust any
	user-installed normal-world 3rd party certificates.
\item The VPN feature in Android allows an application to register as a VPN provider and route all traffic through it. This involves asking a permission 
	upon installation and VPN connection startup.
\end{enumerate}

The attack scenario is an ``Evil Maid'' attack (short-term physical access) against an unlocked device (for example the attacker may 
ask the victim to make a quick phone call from her device). The attack is performed as follows:
\begin{enumerate} \setlength{\itemsep}{0pt}
  \item Install the malicious application requiring VPN-related permissions.
  \item Install a 3rd party certificate.
  \item Run the malicious application which starts a VPN connection.
  This will cause a notification to appear with the icon of the malicious application and name of the VPN connection. 
  \item Serve forged SSL/TLS certificates while performing MITM.
\end{enumerate}

The notification ``red flag'' can be easily mitigated by the attacker. 
By using the KNOX icon and a benign name for the VPN connection such as ``KNOX Connectivity'' 
a non-tech-savvy user will assume that the notification is legitimate part of KNOX and continue using the device normally.

For as long as the VPN connection is active, all device traffic outside and inside KNOX will be routed through the VPN connection.
During this time, in order to intercept SSL/TLS traffic, the malicious application will serve fake website certificates signed with its previously installed 3rd-party certificate.
Due to the shared certificate store, any KNOX application relying on a chain-of-trust verification will believe the certificate to be authentic and
continue operating as normal, allowing the user to disclose her secret data to the attacker. 
Alternatively, the attacker can perform website forgery by serving locally stored copies of sites to the user, thus avoiding the need to serve as MITM
in the communication with the real websites.
This attack will not persist after reboot as there is a need to manually start the VPN connection.

The sole requirement for this attack is the ability to manually install an application in the outside Android environment and click away all the warning 
dialogs. The attack doesn't involve any exploits or require knowing the user's KNOX password. 
It should be noted that the attack was not tested in conjunction with KNOX's enterprise per-app-vpn feature. 

This critical vulnerability demonstrates the danger in sharing resources (in this case, the certificate store) between the user environment and the secure container. 
Whenever possible, the secure container should use separate resources and if that is not possible, then verify all shared resources prior to usage.

In response to our vulnerability report \cite{SamsungCorrespondence}, Samsung informed us that the vulnerability was already known to them as a limitation of KNOX 1.0 
and was corrected in KNOX 2.0. 

Another different in nature yet VPN-related attack involving capturing of traffic on Android was demonstrated by BGU researchers in \cite{BGUVpnRelatedAttack}. 
Whereas this attack redirects traffic away from a VPN connection, ours uses the VPN connection to redirect the traffic. 
The novel contributions of our attack are redirection of traffic from KNOX as well as from the user environment and the ability to 
perform a MITM attack on SSL/TLS traffic.

\subsubsection{Clipboard} \label{knox-1.0-attacks-clipboard}
In this section we present CVE-2016-3996 \cite{CVE-2016-3996}, a vulnerability that allows an attacker to steal the contents of the KNOX clipboard.
One of the KNOX proprietary services is \emph{clipboardEx}. It provides access to both the KNOX clipboard and the Android clipboard.
KNOX applications that wish to access the clipboard use the client class \emph{ClipboardExManager} to connect to the 
service that runs in the \emph{system\textunderscore server} and is implemented by \emph{InternalClipboardExService}.
The ClipboardExManager is not available through the standard SDK for Android developers but is preloaded by \emph{Zygote} into every application process 
from \emph{framework2.jar/odex} under \emph{/system/framework}.

The \emph{InternalClipboardExService} can only represent one ``clipboard" at a time. The choice is determined by 
an internal member named \emph{mContainerID}. When \emph{mContainerID} is set to zero the service works with the user environment and 
any other value indicates the KNOX clipboard.
An important method of the service is \emph{UpdateClipboardDB(int containerId)}. Calling the method updates the mContainerID member accordingly, 
without checking who is the caller, i.e., anyone that knows about the service can bind to it and call this method. 

An additional interesting fact is that in Android the clipboard data is persistent and is stored under 
\emph{/data/clipboard} (user) and \emph{/data/clipboard/knox} (KNOX) accessible only by the \emph{system} user.
Both the user and the KNOX clipboard data is stored unencrypted, outside the eCryptFS file system. \\

The client-side \emph{ClipboardExManager} provides the API to query the \emph{clipboardEx} service.
In the beginning of each method the ClipboardExManager obtains an instance of \emph{IClipboardService} and uses it to perform the API call. 
\emph{IClipboardService} is the AIDL-generated binder wrapper for the \emph{clipboardEx} API.

However, before each call to the \emph{IClipboardService} method, \emph{ClipboardExManager} calls \emph{checkCurrentMode()} which 
among other things calls \emph{UpdateClipboardDB(0)} from \emph{IClipboardService} (when called from the normal world), setting the mContainerId 
of the server to 0, thus preventing access to the KNOX clipboard. \\

In order for an attacker to get access to the KNOX clipboard all that is needed is for the service to have mContainerID != 0.
The following issues stand in an attacker's way:
\begin{itemize} \setlength{\itemsep}{0pt}
\item The \emph{ClipboardExManager} is not in the SDK so it is harder to use.
\item \emph{UpdateClipboardDB} isn't exported from \emph{ClipboardExManager}.
\item Each API call via ClipboardExManager performs \emph{checkCurrentMode()} to reset the service to the appropriate container ID.
\end{itemize} 

\noindent These protections can be easily bypassed by:
\begin{itemize} \setlength{\itemsep}{0pt}
\item Using framework2 from a device or firmware.
\item Calling the private static method \emph{ClipboardExManager.getService} via reflection to get an \emph{IClipboardService} instance.
\item Using the APIs in \emph{IClipboardService} directly instead of the wrapper methods in \emph{ClipboardExManager} to avoid the calls to checkCurrentMode().
\end{itemize}

The code exploiting the vulnerability, depicted in Listing \ref{lst:ClipboardAttack}, works on 
KNOX 1.0 regardless of whether KNOX is in the foreground or background or even unlocked.

\begin{lstlisting}[breaklines=true, caption={Clipboard Attack Code}, label={lst:ClipboardAttack}]
Method getService = ClipboardExManager.class.getDeclaredMethod("getService");
getService.setAccessible(true);
IClipboardService clipService = (IClipboardService)getService.invoke(null);
clipboardService.UpdateClipboardDB(1);
int size = clipService.getDataSize();
ClipboardData data = clipService.GetClipboardData(1);
ArrayList<String> clips = clipService.getClipedStrings(0, size);
\end{lstlisting}

The primary implication of this vulnerability is that any application without any permissions can get access to the KNOX clipboard.
Given that copy-paste is an invaluable tool, this vulnerability can lead to a large scale disclosure of sensitive data.
A primary security principle in Android services is that the service must always validate that the caller has the appropriate permissions to perform
the requested action (either by checking the UID and PID or by checking that the application has some Android permission). 
Although in this case the attack was made possible owing to the \emph{clipboardEx} service not following this principle, the attack does highlight a crucial 
``secure container'' design pitfall; special attention must be paid to securing any resource used by KNOX and accessible to the user (in this case the service).

Moreover, although insignificant compared to the primary vulnerability, the fact that the KNOX clipboard is not encrypted, unlike the rest
of the KNOX data, is a security hole in itself.
It means that a root or system user can simply read the persistent clipboard data without having to know the user's password.

This clipboard vulnerability was unknown to Samsung when we reported it. Samsung's response was that users should upgrade to 
KNOX 2.3 \cite{SamsungCorrespondence}. This version, however, is still vulnerable to a modified version of this attack as we show in section \ref{knox-2.x-attack-clipboard}.

\subsubsection{ADB} \label{knox-1.0-attacks-adb}
In this section we describe how an attacker can abuse the \emph{Android Debug Bridge} (ADB) development feature to attack KNOX.
\emph{ADB} consists of a PC client, and the \emph{adbd} daemon on the device (running as the \emph{shell} user), 
whose purpose is to execute commands from the client.
This feature is disabled by default and has to be manually enabled by the user via the UI.

The isolation between applications has been mostly implemented well using SELinux and various security checks in central services 
(ActivityManager, PackageManager).
This isolation prevents user applications from communicating with services running inside KNOX applications, sending broadcasts, querying content 
providers and starting activities.
However, we have detected a loophole allowing the \emph{shell} user to launch KNOX activities and send broadcasts. 
This allows an attacker having a foothold on the user's computer to attack KNOX via USB if it is unlocked (after the user's password has been entered). 

Using \emph{ADB} an attacker can launch the KNOX secure world browser with a target URL leading to a website under the attacker's control. 
See example provided in Listing \ref{lst:BrowserAttack}.

An additional security hole is the ability to send broadcasts that will be received by KNOX applications as well.
The effect can be devastating if an application relies on a broadcast to trigger or disable some feature.
For instance, a broadcast can change the KNOX browser's search engine. 
See example provided in Listing \ref{lst:BroadcastAttack}.

\begin{lstlisting}[breaklines=true, caption={Launching Browser from ADB}, label={lst:BrowserAttack}]
adb shell am start -a android.intent.action.VIEW -n sec_container_1.com.sec.android.app.sbrowser/com.sec.android.app.sbrowser.SBrowserMainActivity -d "http://www.attackerwebsite.com"
\end{lstlisting}

\begin{lstlisting}[breaklines=true, caption={Sending Broadcasts from ADB}, label={lst:BroadcastAttack}]
adb shell am broadcast -a sec_container_1.android.intent.action.CSC_BROWSER_SET_SEARCH_ENGINE --es searchEngine bing
\end{lstlisting}

\subsubsection{Lessons Learned}
\paragraph{Limiting the Attack Surface}
It is very difficult to design a secure container and to account for every possible attack, as demonstrated by the 
above attacks. However, two of them could have been prevented by limiting the secure container's attack surface.
Concretely, the clipboard attack (\ref{knox-1.0-attacks-clipboard}) by not exposing the \emph{clipboardEx} service to non-KNOX applications and 
the ADB attack (\ref{knox-1.0-attacks-adb}) by avoiding communication with processes run as the \emph{shell} user 
or disabling ADB altogether (the solution chosen by KNOX 2).

\paragraph{Avoiding Resource Sharing} \label{knox-1.0-lesson-resource-sharing}
Sharing resources between the secure container and the insecure environment is a recipe for disaster. We've seen this through the sharing of the \emph{connectivity} 
service and the certificate store in the VPN attack \ref{knox-1.0-attacks-vpn-mitm}. In particular, any resource that can be either monitored (e.g., network data) or modified 
(e.g., certificate store, GPS coordinate source) must not be shared between the two environments.

\subsection{Root-Dependent Attacks} \label{knox-1.0-root-attacks}
As we've seen, in KNOX 1.0 the scenario where an attacker with \emph{root} privileges co-exists on a device while KNOX is installed and even running is 
quite possible in the case that TIMA fails to prevent a runtime \emph{root} attack, thus invalidating KNOX's TCB. 
Such a scenario has been confirmed as possible by Samsung in our correspondence \cite{SamsungCorrespondence}, stressing that their aim is to 
limit the harm whenever possible.
In this section we review what a root-privileged attacker can do to exfiltrate or corrupt sensitive KNOX data. 

\subsubsection{Volatile Access to KNOX Data}
In this section we describe root attackers' access to KNOX encrypted data. 
We've mentioned that logging into KNOX mounts the eCryptFS containers for the application data and sdcard onto \emph{/data/data1} and 
\emph{/mnt\textunderscore 1/sdcard\textunderscore 1} respectively. While they are mounted these directories are accessible to the root user and provide 
read and write access to the KNOX data making the encryption underneath completely transparent.

Moreover, these filesystems remain mounted even when the user logs out of KNOX or is auto-locked after a timeout. 
We found that once the eCryptFS containers are mounted they remain mounted until device power-off regardless of the KNOX state, 
exposing the data to a root attacker.
Note that the attacker may choose to modify and / or corrupt the data thereby altering KNOX application functionality or causing a Denial of Service.

\subsubsection{eCryptFs Key} \label{knox-1.0-ecryptfs-key}
In this section we present CVE-2016-1919 \cite{CVE-2016-1919}, a vulnerability that allows an attacker to decrypt KNOX encrypted 
data without knowing the user's password.
In section \ref{knox-1.0-ecryptfs} we described the eCryptFS paradigm in KNOX, but we left out the process of combining the user's
password and the TIMA key to produce the \emph{eCryptFS key}. As specified by the KNOX 1.0 whitepaper \cite{AnOverviewOfSamsungKNOX1}, 
the key used for the encryption is derived by ``well-known key-derivation algorithms such as Password-Based Key Derivation Function 2 (PBKDF2)''.
We did observe the usage of PBKDF when generating the Master Key, but it is rendered meaningless due to the poor generation of the \emph{eCryptFS key}. 
Listing \ref{lst:ecryptfs-key-generation} shows the precise algorithm used to derive the \emph{eCryptFs key}. 

\begin{lstlisting}[float=h, breaklines=true, numbers=left, caption={eCryptFS Key Generation}, label={lst:ecryptfs-key-generation}]
private static String getEcryptFsKey(String password, byte[] timaKey) {
    byte[] bytes = String.format("%1$32s", password).getBytes();
    byte[] keyBytes = new byte[32];
    for (int i = 0; i < 32; i += 1) {
        keyBytes[i] = (byte) (bytes[i] ^ timaKey[i]);
    }
    String ecryptFsKey = Base64.getEncoder().encodeToString(keyBytes);
    return ecryptFsKey.substring(0, 32);
}
\end{lstlisting}

We can partition the algorithm as follows:
\begin{compactdesc}
\item [Line 1] Left-pad the password with spaces to 32 chars
\item [Lines 2-6] Byte-wise XOR the padded password with the TIMA key to produce ``random'' 32 bytes
\item [Line 7] Encode the ``random" 32 bytes in \emph{base64}
\item [Lines 8-9] Return the 32 left-most chars of the base64-encoded string as the \emph{eCryptFS key}
\end{compactdesc}
The problem with this algorithm is that Base64 expands the given input with a ratio of 4:3 (every 3 bytes result in 4 chars).
Given this ratio, only the leftmost 24 bytes of the XOR'ed sequence actually affect the eCryptFS key.
If the password is up to 8 characters long, the user's password will be completely ignored and only the padded spaces will mix with the TIMA key.
Seeing as the minimal length for a user's password is 7 characters, it is reasonable to assume that most users will choose a password of that length. 
Given a somewhat longer password, all it will take is a very simple bruteforce attack.

Moreover, since a root attacker can obtain the TIMA key through the TIMA service, this vulnerability places the \emph{eCryptFS key} and 
subsequently all of the encrypted data firmly in the attackers hands.
Additionally, if the user were to change her password to any other up-to-8-chars sequence, it wouldn't actually change the \emph{eCryptFS key}, 
fooling the user into thinking she has somehow changed the way her data is protected.

We speculate that the reason for this critical flaw is the textual interface when communicating with \emph{vold}, 
requiring the password to contain only printable characters. This probably lead the algorithm's designer to encode the potentially 
unprintable bytes in base64 yet take only the left-most 32 due to the eCryptFS password's length limitation.

This attack demonstrates yet another pitfall, for which encrypted data storage, due to its inherent complexity is a prime candidate. 
We see that integration of several components, each secure on its own (eCryptFs key generation without truncation and the eCryptFs filesystem) can lead 
to an insecure combination with hazardous results. 

In response to our report, Samsung informed us that this vulnerability was identified during an internal security review and was corrected in KNOX 2.0 \cite{SamsungCorrespondence}.

\subsubsection{Data Encryption Key (DEK)} \label{knox-1.0-dek-attack}
In this we section we outline attacks aimed at obtaining the DEK.
As we recall from section \ref{knox-1.0-ecryptfs}, a design flaw in the process of obtaining the DEK is that eventually it resides in \emph{plaintext} 
in \emph{vold}'s memory. In a TrustZone paradigm, one would expect that the DEK would be accessible only to the secure world, protecting it from 
root attackers in the normal world. 

Suppose an attacker with root privileges knew the eCryptFS key at one point in time. 
Then as root he could read the EDK Payload file and could then generate the MK and decrypt the EDK, obtaining the DEK. 
This would give the attacker permanent access to the encrypted data regardless of user password changes 
(because the DEK will always stay the same, only the MK would change).

Looking into the method reading the EDK file (\emph{ssRead}), we discovered that after reading the file, the method decrypts it using 
\emph{SecureStorage} API which is in fact implemented in the TrustZone (via a MobiCore OS trustlet).
This lead us to conjecture that running as an external root process, dynamically loading \emph{libsec\textunderscore ecryptfs.so} and calling \emph{ssRead} 
would simply get us the EDK Payload. This attack failed due to the TrustZone refusing to decrypt the EDK Payload. The reason is unclear and warrants 
further research.

A second attack vector we attempted was injecting code into \emph{vold}'s memory and hooking \emph{ssRead} with the purpose of obtain the EDK 
payload during a legitimate code flow (while mounting a container). 
This attack fails as well as some component in the call chain to the TrustZone detects the hook placement in \emph{vold}'s memory.

These defences have indeed proven effective against the two attacks we tried but they still don't prevent the running of injected code inside \emph{vold} 
while it mounts the containers (which we've successfully accomplished). 
This code could still, at the opportune moment, pull the decrypted EDK payload or even the DEK from \emph{vold}'s memory.
Thus we hypothesize that an attack to extract the permanent EDK is, in principle, viable.

\subsubsection{Keyboard Sniffing} \label{knox-1.0-keyboard-sniffing}
In this section we describe how a root attacker can sniff the keyboard input to KNOX using Java code injection (see section \ref{knox-1.0-architecture}).
As we already mentioned, Samsung has taken a precaution against malicious keyboard applications by limiting KNOX to work only with the Samsung 
official keyboard. This, however, doesn't defend against root-enabled attackers.
The ``input method'' architecture in Android works as follows:
\begin{itemize}[itemsep=-2pt,topsep=0pt,leftmargin=1em]
\item Each input method (e.g., keyboard) runs in its own application process
\item Each application requiring user input runs its own process as well
\item The mediator is the \emph{input\textunderscore method} service running in the \emph{system\textunderscore server}.
\end{itemize}
This design means that the keystroke data passes through each of the processes during processing. KNOX applications use the same service and 
subsequently the same keyboard process as normal user applications (running outside KNOX). This means that each of them, compromised by an attacker
can lead to disclosure or alteration of keystroke data. This scenario obviously applies to root attacks but also to exploitable vulnerabilities in the 
Samsung keyboard like \cite{CVE-2015-4640}. Following this attack vector, we have successfully implemented a fully functional KNOX keyboard sniffer. 
Using the \emph{Xposed} framework we injected Java code to the Samsung keyboard process and inserted hooks into the code flow processing the keystrokes. 

\subsubsection{Screen Capture} \label{knox-1.0-screen-capture}
In this section we describe how a root attacker can take screenshots inside KNOX using Java code injection (see section \ref{knox-1.0-architecture}).
Android uses the concept of secure windows to disable screenshot functionality (even when manually pressed).
KNOX applications use this feature to protect themselves from being spied on. However this protection can be disabled
by a root attacker which can disable the ``secure flag'' on windows running inside KNOX processes and by doing so expose
them to screenshots (e.g., by running \emph{/system/bin/screencap}). We have successfully implemented this attack by injecting code into all KNOX processes 
using the \emph{Xposed} framework and preventing the ``secure flag'' from being turned on. This allowed us to freely take screenshots during the 
KNOX login process and inside KNOX without alerting TIMA.

\subsubsection{Hiding the Warranty Bit from KNOX} \label{knox-1.0-hide-warranty-bit}
In this section we describe how a root attacker can ``blind'' KNOX and prevent it from detecting that the warranty bit has been set. 
This attack is a concrete example of the danger that Java code injection poses to KNOX, as we've observed in section \ref{knox-1.0-architecture}.
We performed this attack on the S3 which has been rooted by flashing a custom firmware, subsequently setting its warranty bit.
We revealed in section \ref{knox-1.0-tima-key} that the TIMA key is generated in user-mode and is ``installed'' and ``retrieved'' via the TIMA service.
Moreover, in section \ref{knox-1.0-tima} we've uncovered that the TIMA key ``installation'' API call is the one that prevents us from running KNOX
on a device whose warranty bit has been turned on. Combining the two facts leads to a solution allowing to bypass the warranty bit check
and to obtain a fully functional KNOX.

Using the \emph{Xposed} framework we injected code to the \emph{system\textunderscore server}. 
By overriding the \emph{keystoreInstallKey} to always succeed and \emph{keystoreRetrieveKey} to consistently return the same key, we avoided communicating
with the TIMA Keystore and supplied KNOX the TIMA key needed to continue creating the container, fully ``enabling'' KNOX on our rooted Galaxy S3.  
The success of this attack indicates that in fact no other KNOX APIs test the warranty bit, in particular the SecureStorage API which is crucial to 
mounting the encrypted file system.

Note that this attack cannot be used to ``re-enable'' an already active KNOX container after rooting due to
inability to obtain the original TIMA key. However, it can be used by an attacker having access to KNOX-designated devices before they reach the end-user. 
The attacker can root the device and insert a root-privileged backdoor capable of fooling KNOX and launching a wide variety of attacks while running 
alongside it.

\subsubsection{Lessons Learned}
\paragraph{Data Encryption}
The attacks presented in this section involving the data encryption scheme have lead to the conclusions described in section \ref{knox-1.0-data-encryption-lesson}. 

\paragraph{Periodic Detection}
We have shown that if a malicious root attacker can get through KNOX's defences he can extract a plethora of sensitive information 
as well as corrupt the system. The lesson to be learned is that most of the efforts must be placed on not letting the attacker through to begin-with 
(which is in fact, KNOX's approach). 
On the other hand, the secure container should use the TrustZone (assuming it at least is not compromised), to discover such attackers whenever possible.
The situation depicted in the attack in section \ref{knox-1.0-hide-warranty-bit} could have been avoided by checking the warranty bit in additional API 
calls (e.g., \emph{SecureStorage}) or even periodically via \emph{PKM}.

\section{KNOX 2.X and Beyond} \label{knox-2.x}
After having thoroughly examined KNOX 1.0, in this section we review KNOX 2.3, the most recent of KNOX to date \cite{SamsungKNOX2Announcement}. 
To conduct our research we used a Note 3 Model SM-N9005 running Android 5.0 and KNOX Version 2.3 (Kernel version 3.4.0, build number LRX21V.N90055XXUGBOJ6).

\subsection{Changes from KNOX 1.0}
Samsung released the latest version of KNOX \cite{AnOverviewOfSamsungKNOX2} after the release of Android 4.4 relying heavily on its features.
Like in KNOX 1.0, KNOX 2.x is a framework, embedded into the device starting from the boot process, through the TrustZone and up to the Android Application Framework along with a 
front-end in the form of an Android application.
Unlike KNOX 1.0, which only had a single front-end application pre-deployed on the device, KNOX 2.3's front-end application comes in multiple flavours. 
Besides the default pre-deployed version there is ``My Knox'' available for download from Google Play
and there are others, each version supporting additional features mainly geared towards the enterprise clients.

The concept of ``multiple users'' was introduced on shareable devices (e.g., tablets) in Android 4.2 with full support for all devices only since Android 5.0.
Samsung, requiring the feature for KNOX 2.0 prior to Android 5.0, enabled it only on their smartphones starting from Android 4.4.
This feature allows multiple people to use the same physical device, providing each one a separate environment 
with their own applications. Applications of multiple users can run in the background, while only one user environment is in the foreground. 
Users are mostly separated from each other but controlled data sharing is supported.
Google utilizes this feature for its own ``Android for Work'' \cite{AndroidForWorkSecurityWhitePaper} security solution by making the work environment 
simply run as another user relying on the ``user separation'' for isolating the work container from the user environment.
KNOX 2.3 does the very same thing, running the applications of the KNOX container as a separate user.

KNOX 2.3 suffers from the same design flaw as KNOX 1.0, running all user and KNOX applications in the same ``Android'' environment, 
side by side, still all forked from \emph{Zygote} and still sharing the same \emph{system\textunderscore server}.
As in KNOX 1.0, the isolation between (multiple) users and the KNOX container relies on SELinux. 
In KNOX 1.0 user applications and KNOX applications run under different SELinux contexts, relying on SELinux policy to thoroughly define the boundaries 
between the two. Instead in KNOX 2.3 all applications run under the same context (\emph{untrusted\textunderscore app}) but under different SELinux 
categories which take care of the isolation. This change simplifies the creation of the SELinux policy and leaves less room for errors.

As noted in section \ref{knox-1.0-app-wrapping}, KNOX 1.0 has a mandatory app-wrapping requirement for installing applications inside KNOX due to 
package name collisions.
This restriction was resolved in Android 4.4 by the ``users'' feature, clearing the way for KNOX 2.x to allow installation of any
application inside the container without the need for wrapping. 
With the technical limitation lifted, Samsung also made the business decision to no longer force KNOX applications to originate in the Samsung Store; 
KNOX 2.x allows installing applications from Google Play and other sources.
Although this is a leap forward in both productivity and usability of KNOX, it is a setback in terms of security: 
in KNOX 1.0 Samsung was responsible for what the user could install in her container, serving as an effective malware filter. 
Now the responsibility has been passed over to the end-user and the IT administrators in the enterprises.
In section \ref{knox-2.x-attack-data-exfil} we show some of the practical repercussions of this decision.

On the other hand, KNOX 2.3 solves many of the security issues we saw in KNOX 1.0.
The ``eCryptFS Key'' vulnerability was corrected as the password management and encryption were thoroughly revised making much more extensive usage 
of the TrustZone.
The ``VPN MITM'' attack is also no longer applicable as KNOX 2.3 supports a separate certificate store for the KNOX container as well as separate 
VPN routing. Moreover, the ADB debugging features were disabled altogether while KNOX is installed on the device. This very effectively mitigates any
PC-to-device attacks like the ones we've shown. An additional security enhancement comes in the form of separate ``keyboard'' processes for each user, 
making it more difficult for attacks on the user keyboard to affect KNOX.

\subsection{Attacks} \label{knox-2.x-attacks}
\subsubsection{Clipboard} \label{knox-2.x-attack-clipboard}
In this section we continue describing CVE-2016-3996 \cite{CVE-2016-3996}.
KNOX 2.3 adds a new feature in the form of controlled clipboard sharing. The user may selectively choose to share certain clips between the Android 
environment and KNOX.
This feature is controlled via a policy that can be set by the IT administrators.
However, the service architecture still relies on a shared \emph{system\textunderscore server} process, providing access to the 
services running within to all users. Apparently in the course of redesigning \emph{clipboardEx} to support clipboard sharing, some security measures 
were added. When asked to provide clipboard data the service checks against clipboard sharing policy, whether KNOX is active at the moment and which user made 
the request. 

These changes effectively mitigate the ``simple" clipboard attack we showed in section \ref{knox-1.0-attacks-clipboard}. 
When the attacker requests the clipboard data, \emph{clipboardEx} either returns the user clipboard or refuses to relinquish the KNOX data.

Unfortunately these security improvements aren't enough. Other APIs, including \emph{UpdateClipboardDB} and querying the clipboard size remain unprotected. 
We were able to reproduce our attack by creating a new user activity while KNOX is running in the background. 
We implemented the attack by installing two ``benign'' applications in the Android environment. 
The first application's role was to launch a service that waits for the user to log into KNOX. At this time the service launches an activity from the 
second application, which in its \emph{onCreate} method, executes the code depicted in Listing \ref{lst:ClipboardAttack} and calls \emph{finish()}, 
thus it appears only for a split-second on the screen. 
The \emph{activity} launch triggers a race condition in the \emph{clipboardEx} service allowing us to temporarily (for several seconds)
obtain the contents of the KNOX clipboard. The first application's service is capable of repeating this attack for as long as KNOX remains in the foreground.

This attack was successfully tested on the KNOX 2.3 application bundled with the Note 3 as well as on ``My Knox'' which can be downloaded from Google Play.
Merely updating ``My Knox" on Google Play is not enough to patch this vulnerability, since it lies within the \emph{system\textunderscore server}, 
which is part of the core system image, thus requiring a full system update.

Following our report in December 2015, Samsung were able to reproduce and isolate the race condition responsible for the vulnerability and 
starting distributing a patch around March 2016 \cite{SamsungCorrespondence}.

\subsubsection{Data Exfiltration} \label{knox-2.x-attack-data-exfil}
Much of Android's protection of application data relies on the permission model. Each application requests a set of permissions to perform the actions it
requires. The user can accept or reject these permission requests upon application installation: the typical usual user action is to accept them 
and continue with the installation.
In KNOX 2.3 the user may install an application from Google Play or transfer an application already installed 
in the Android environment to KNOX. This feature exposes the sensitive data within the KNOX container to dangers from
malware by abusing the Android permission mechanism and relying on the blind user acceptance. 
Note that installing applications inside KNOX does not require root privileges - any user can do it.
The only obstacle standing in an attacker's way is the vigilance of the IT administrators, which must effectively use the MDM 
application white- and black-listing capabilities. In the case of a non-IT managed KNOX container, the responsibility to watch for 
malicious applications lies directly on the user. 

To demonstrate these dangers we've created a ``backdoor'' application, that when installed inside KNOX is capable of: 
communicating with an Internet C\&C server, downloading extension modules, uploading data exfiltrated from within KNOX,
and sending SMS messages.
Our information-stealing app is able to extract KNOX private contacts and calendar details, KNOX clipboard contents, user-environment SMS messages,
KNOX browsing and search history, and sdcard files such as KNOX photos and downloads.

It is important to note that in order to slip under the radar a malicious application can initially behave in a completely benign manner (without even 
containing any malicious code), yet request all permissions required for the malicious behaviour. Once it is installed within KNOX, it can update itself to 
its malicious version without triggering the user's suspicion as it won't even request any new permissions. This implies that IT admins must
validate applications upon each update and not only upon initial installation. A possible mitigation for this scenario is DroidDisintegrator \cite{DroidDisintegrator}, 
a system for characterizing information flows within applications, presenting them for users' approval, and ensuring that application updates do 
not introduce new and unapproved information flows. A similar approach for preventing sensitive data leakage is used in TrustDroid \cite{TrustDroid}.

This demonstrates that applications within the KNOX container are not strongly segregated from each other and IT administrators need to be constantly on
the lookout for dangerous applications and use the MDM APIs to monitor and protect their company's devices. 

\subsubsection{Lessons Learned}
\paragraph{Avoiding Resource Sharing}
The presence of the clipboard attack \ref{knox-2.x-attack-clipboard} in KNOX 2.x, even in light of the added security features, only strengths our 
argument for avoiding sharing of resources between the secure and insecure environments (see section \ref{knox-1.0-lesson-resource-sharing}).

\paragraph{Data Protection}
Special care must be taken to protect the data within the container, not only from outside but also from within. 
Verification of applications prior to installation by a trusted party can help reduce the risk but not eliminate it entirely. 
Moreover, and especially in the case that such verification doesn't take place, to better handle malware finding its way into the secure container, 
we recommend deploying a malware and data leakage detector within the secure container.

\section{Conclusion} \label{conclusion} 
We have presented an extensive security assessment of critical security features in the paradigm of secure containers for Android.
Each aspect was demonstrated through a real-world example of a security solution deployed on millions of devices worldwide.
Our research has revealed the inner workings on KNOX, contrasting the vendor's security claims with reality. We identified several design weaknesses 
and some actual vulnerabilities.
Through our analysis we presented concrete lessons and guidelines for designing and implementing a secure container.
We highlighted the danger of sharing KNOX services with user applications, even in the presence of dedicated security measures. 
The sharing of services is a two-edged sword: on the one hand allowing simpler design and implementation, but on the other hand creating a constant 
security threat. 
We demonstrated the dangers that root and kernel exploits present and the importance of properly mitigating them through a hardware root of 
trust supported by the ARM TrustZone.
We also showed that the TrustZone's mere existence is not enough, requiring proper usage of its features in all surrounding areas to gain the promised security boost.
We pointed out the dangers posed by simple applications to information within the secure container as well as the importance of 
closely tracking application updates.
Our findings were shared and discussed in details with the vendor, allowing sufficient time for patches to be distributed.
We hope that our work will help future designers avoid potential pitfalls by highlighting crucial areas, improving future BYOD security solutions.

\subsection*{Acknowledgements}
We would like to thank Dr. Eran Tromer for his valuable input that helped us improve the quality of this paper.
We also like to extend our deepest appreciation to the Samsung Mobile Security Team for their responsiveness and cooperation in the process
of disclosing the detected vulnerabilities as well as for providing us with better insight into the design of Samsung KNOX making this paper better.

{
\footnotesize 
\bibliographystyle{acm}
\sloppy
\bibliography{Tex/Bibl}
}

\end{document}